\documentclass[reprint,amsmath,amssymb,aps]{revtex4-2}

\usepackage{graphicx}
\usepackage{dcolumn}
\usepackage{bm}
\usepackage[driverfallback=dvipdfmx,colorlinks,linkcolor=blue,citecolor=blue,urlcolor=blue]{hyperref}
\usepackage[mathlines]{lineno}

\begin{document}

\title{Interference of the scattered vector light fields from two optically levitated nanoparticles}

\author{Yuanbin Jin}
\affiliation{State Key Laboratory of Quantum Optics and Quantum Optics Devices, Institute of Opto-Electronics, Synergetic Innovation Center of Extreme Optics, Shanxi University, Taiyuan 030006, P.R.China}
\author{Jiangwei Yan}
\affiliation{State Key Laboratory of Quantum Optics and Quantum Optics Devices, Institute of Opto-Electronics, Synergetic Innovation Center of Extreme Optics, Shanxi University, Taiyuan 030006, P.R.China}
\author{Shah Jee Rahman}
\affiliation{State Key Laboratory of Quantum Optics and Quantum Optics Devices, Institute of Opto-Electronics, Synergetic Innovation Center of Extreme Optics, Shanxi University, Taiyuan 030006, P.R.China}
\author{Xudong Yu}
\affiliation{State Key Laboratory of Quantum Optics and Quantum Optics Devices, Institute of Opto-Electronics, Synergetic Innovation Center of Extreme Optics, Shanxi University, Taiyuan 030006, P.R.China}
\author{Jing Zhang}
\email{jzhang74@sxu.edu.cn, jzhang74@yahoo.com}
\affiliation{State Key Laboratory of Quantum Optics and Quantum Optics Devices, Institute of Opto-Electronics, Synergetic Innovation Center of Extreme Optics, Shanxi University, Taiyuan 030006, P.R.China}

\begin{abstract}
We experimentally study the interference of dipole scattered light from two optically levitated nanoparticles in vacuum, which present an environment free of particle-substrate interactions. We illuminate the two trapped nanoparticles with a linearly polarized probe beam orthogonal to the propagation of the trapping laser beams. The scattered light from the nanoparticles are collected by a high numerical aperture (NA) objective lens and imaged. The interference fringes from the scattered vector light for the different dipole orientations in image and Fourier space are observed. Especially, the interference fringes of two scattered light fields with polarization vortex show the $\pi$ shift of the interference fringes between inside and outside the center region of the two nanoparticles in the image space. As far as we know, this is the first experimental observation of the interference of scattered vector light fields from two dipoles in free space. This work also provides a simple and direct method to determine the spatial scales between optically levitated nanoparticles by the interference fringes.
\end{abstract}

\maketitle

\section{Introduction}
Interference plays an important role in exploring the nature of light, i.e., the wave-particle duality \cite{Wootters79,Englert96} and has extensive applications such as in precise measurement \cite{Millen14,Zhang2017,Hebestreit2018,Ahn20}. Interference of light scattered from particles has received much attention because it has provided information on the degree of localization of particles in the trap \cite{Grangier85,Eichmann93,Wolf16,Araneda18}. As the techniques for trapping atoms and ions developed, the interference of faint scattered light from the atoms and ions have been reported \cite{Grangier85,Eichmann93,Wolf16,Araneda18}. The previous studies usually considered the scalar fields (or fields with the same polarization). In recent years, the coherence of vector light fields in quantum and classical regimes have been a new focus \cite{Gori06,Eberly17,Norrman17,Qian20,Norrman20}, in which the polarization is taken into account in the coherence theorem and complementarity principle. The Young’s double-slit interference illuminated by a vector field has been explored \cite{Grangier85}.

In recent years, the optically levitated micro- and nano-particles in vacuum have attracted much attention and became an important platform for the precise measurement and fundamental physics studies \cite{Li10,Aspelmeyer14,Arndt14,Gieseler14,Millen14,Ranjit15,Jain16,Hoang18}. Cooling the center-of-mass motion of the trapped nanoparticle to reach quantum ground state has been realized recently \cite{Deli20,Magrini21,Tebbenjohanns21}. The optically levitated nanoparticle has been used to measure the Casimir force \cite{Xu17}, torque \cite{Ahn18,Ahn20} and realize GHz scale rotation \cite{Ahn18,Ahn20,Reimann18,Jin2021}. In this system, the micro- and nano- particles are free from particle-substrate compared to other systems which have particle-environment interactions \cite{Sick00,Monica05,Dyla17,Lernereaan1133,Zhou19}. Recently our group reported the imaging of the dipole scattering orientations of an optically levitated silica nanoparticle in the image and Fourier space (k-space) \cite{Jin21}. This system has several distinguished features. First, the optically levitated nanoparticle provides the possibility to be illuminated with an excitation laser beam orthogonally to the optical axis of objective lens. Thus this scheme provides a dark background and high signal-noise ratio for detecting the dipole scattering. Secondly, vector light field has been known in the context of classical optics, such as conical diffraction \cite{Berry200713} and propagation through anisotropic crystals \cite{Oron00}, singular optics \cite{Soskin2001219}, spin–orbit coupling \cite{Cardano2015}, and more recently quantum states \cite{Spreeuw1998,Rosales2018}. Vector light field refers to the light field with spatially varying polarization and in particular polarization vortex beam refers to the light field which reveal an azimuthal change of polarization orientation in its transverse plane. In this dipole scattering system, the scattered vector light fields can be produced, and in particular polarization vortex field with radial polarization is created when the dipole orientation of nanoparticle is aligned exactly along the optical axis of objective lens. Consequently, this system provides an important platform for studying the dynamical properties and the scattering anisotropy. Here, the imaging of dipole scattering orientations of an optically levitated nanoparticle is similar to 3D dipole orientations of a single-molecule, which have been extensively studied by the polarization analysis of emitted fluorescence \cite{Fourkas01,Zhanghao19,Lethiec14}, or by the recording of the defocused \cite{Jerzy97,Bartko99,Karedla15} or aberrated \cite{Bartko992} or k-space \cite{Lieb04,Dodson14,Osorio15,Curcio20} fluorescence image of a molecule. Compared with the single molecule image in which the excitation light is illuminated confocally along the optical axis of objective lens, in our work the illuminating light on the nanoparticle is perpendicular to the optical axis of objective lens.

In this paper, we experimentally investigate the interference of the scattered vector light fields from two optically levitated nanoparticles by employing the distinct advantage of this system. The two nanoparticles are trapped in vacuum using two 1064 nm laser beams with the frequency difference of 110 MHz and orthogonal polarization, strongly focused by a high NA objective lens. We shine a 532 nm laser beam on the two trapped nanoparticles in the direction perpendicular to the propagation of the trapping beams. The scattered 532 nm light is collected by the same high NA objective lens which is used to trap nanoparticles, and imaged on a CCD. The interference fringes of the scattered light in the image and Fourier space are observed. We experimentally investigate the interference of the scattered vector light fields produced for the special dipole orientation from two free-space dipoles and show the distinct interference pattern in intensity and polarization distributions. The interference fringes are measured through an imaging system and can be used as a precise ruler to calibrate the magnification of the image system, so determine the distance between the two nanoparticles.

\section{Experimental setup and theoretical analyzation}

\subsection{Experimental setup}
\begin{figure}
\centering
\includegraphics[width=3.2in]{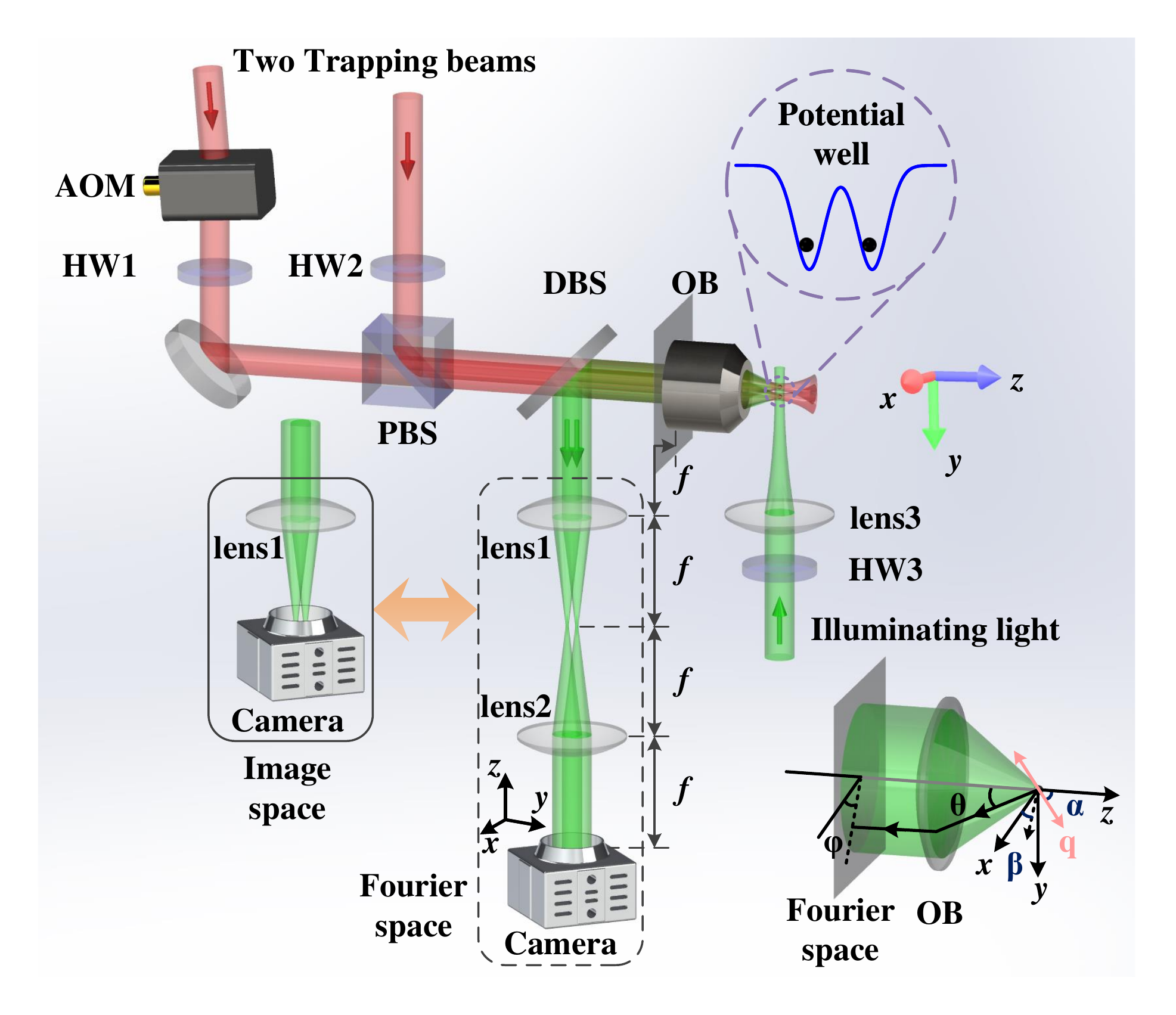}
\caption{Schematic diagram of the experimental setup and the imaging system of dipole scattering. The two trapping beams with slight misalignment are strongly focused by a high NA objective lens (NA=0.95) to simultaneously trap two silica nanoparticles. The 532 nm laser beam is used to illuminate the nanoparticles along the orthogonal direction of the trapped beams ($z$-axis). The scattered light is collected by the same objective lens into a camera. The dipolar scattering is measured in the image space (solid box) and Fourier space (dashed box), respectively. The inset is the diagram of spatial angles of the scattered light. AOM: acousto-optic modulator; PBS: polarization beam splitter; DBS: dichroic beam splitter; HW1-3: half-wave plate. OB: objective lens.}
\label{fig1}
\end{figure}

Fig. 1 represents the schematic diagram of the trapping nanoparticles and the imaging system for measuring the scattered light by the nanoparticles. We have used two 1064 nm laser beams from a diode-pumped single-frequency laser in the experiment. The frequency of one beam is shifted by a mount of 110 MHz using an acousto-optic modulator (AOM), and then it is combined with the other laser beam (the frequency is not shifted) on a PBS to avoid the interference of the two trapping beams. Then the two beams are tightly focused by a high NA objective lens (Nikon CF Plan 100X, NA=0.95) in the vacuum system \cite{Jin18,Jin19}. The directivity of one beam is precisely controlled by a motor-driving mirror. Thus, the two focus points can be adjusted arbitrarily in the radial direction and form two potential wells for trapping two nanoparticles. Simultaneously, the distance of two wells in the axial direction can be changed by adjusting the positions of a collimating lens of the trapping beams. At last, we precisely control the positions of the double wells and keep them on the same object plane of the objective lens.

In the experiment, we utilize silica nanoparticles (Tianjing BaseLine, Unibead silica nanoparticle, diameters are around 100 nm, refractive index = 1.4), which have low dispersion. The commercially available hydrosoluble silica nanoparticles are first dissolved in ultra-pure anhydrous ethanol and then sonicated for 15 minutes. Next the suspension is diluted and poured into an ultrasonic nebulizer. Finally, the droplets containing nanoparticles are dispersed through a narrow tube into the vacuum chamber by the nebulizer. In order to simultaneously load two nanoparticles, in the beginning we have kept the separation between the two wells about 5 $\mu$m. When a nanoparticle gets trapped by one potential well, then we need to slow down the pouring speed to avoid losing the trapped nanoparticle due to air flow. Once two nanoparticles are trapped in the two wells separately, we turn on a vacuum pump to evacuate the chamber. Here, we choose the proper distance (larger than 1.5 $\mu$m) of the two wells in the radial direction and a proper trapping power (both 150 mW) to avoid the nanoparticles jumping forth and back between the two wells. We can stably trap nanoparticle and measured at the pressure from 1 to $10^{5}$ Pa. Here, all measurements are made at the pressure of 1 kPa in the vacuum chamber.

A linearly polarized 532 nm laser beam focused by a lens ($f = 175$ mm) illuminates the trapped nanoparticles. The direction of linear polarization of the illuminating light can be adjusted by a half wave-plate (HW3) before the vacuum chamber. The propagation direction of illuminating beam is along the $y$-axis orthogonal to the trapping beams direction ($z$-axis). The waist of the illuminating beam is about 50 $\mu m$ at the nanoparticle position which is much larger than the distance between the two nanoparticles. The power of the illuminating beam is about 20 $\mu W$, which is weak enough to neglect the influence on the motions of the nanoparticles. The size of the nanoparticles (100 nm diameter) is much smaller than the wavelength of the illuminating light (532 nm). Therefore the two nanoparticles can be approximated as two dipoles. The scattered light from the nanoparticles is collected by the same objective lens, which is used to trap the nanoparticles, and reflected by a dichroic mirror (which is high reflective for 532 nm and anti-reflective for 1064 nm) and imaged on a high resolution CCD camera (Andor Zyla 4.2 sCOMS, 6.5$\mu m$ pixels, 2048 $\times$ 2048).

\subsection{Theoretical analyzation}
We first consider the case of a single nanoparticle briefly. We approximate the nanoparticles as dipoles due to their small size (100 nm diameter) in relation to the wavelength (532 nm) of the illuminating light. In Cartesian coordinates, the dipole orientation induced by the linear polarization of the incident laser is denoted by ($\alpha$, $\beta$), in which $\alpha$ is the zenith angle relative to $z$-axis and $\beta$ is the azimuthal angle in $xy$-plane relative to $x$-axis. Similarly, the k-vector of the scattered field is denoted by ($\theta$, $\varphi$). The intensity distribution in the backaperture plane of the objective lens, which is the Fourier space (k-space) distribution of the scattered light, can be written as \cite{Lieb04,Jin21}
\begin{eqnarray}
{I^F}\left( {\theta ,\varphi ;\alpha ,\beta } \right) \propto \frac{1}{{\cos \theta }}\left( {{{\left| {{\bf{q}} \cdot {{\bf{e}}_p}} \right|}^2} + {{\left| {{\bf{q}} \cdot {{\bf{e}}_s}} \right|}^2}} \right)
\label{eq1},
\end{eqnarray}
where $\bf{q}$ is the dipole orientation unit vector, i.e. ${\bf{q}} = \left( {\sin \alpha \cos \beta ,\sin \alpha \sin \beta ,\cos \alpha } \right)$. The two mutual orthogonal unit electric field vectors ${\bf{e}}_p$ and ${\bf{e}}_s$, which are all orthogonal to the wave k-vector of the scattered field, are defined as ${{\bf{e}}_p} = \left( {\cos \theta \cos \varphi ,\cos \theta \sin \varphi ,\sin \theta } \right)$, ${{\bf{e}}_s} = \left( {\sin \varphi , - \cos \varphi ,0} \right)$.  The NA of the objective lens determines the range of angle $\theta$. The polarization direction of the scattered light field in the Fourier space can be written as ${\bf{p}} = ({\bf{q}} \cdot {{\bf{e}}_p}\cos \varphi  + {\bf{q}} \cdot {{\bf{e}}_s}\sin \varphi ,{\bf{q}} \cdot {{\bf{e}}_p}\sin \varphi  - {\bf{q}} \cdot {{\bf{e}}_s}\cos \varphi )$, which is spatial-dependent and corresponds to the vector light field.

In order to measure the distribution in image space, the scattered light collected by the objective lens is focused through a lens and measured by a CCD camera in the focal region. The distribution of the scattered light in the plane of the dipole is non-paraxial regime of light, however, the observed distribution in image space is the paraxial light of the dipole determined by NA of objective lens. The intensity distribution in the focus region of the lens, which corresponds to the distribution of the scattered light in the image space, is given by the z component of the Poynting vector, \cite{Enderlein00,Bohmer03}
\begin{eqnarray}
{I^I} = \frac{c}{{8\pi }}{{\bf{e}}_z} \cdot \left( {{\bf{E}} \times {\bf{B}}^*} \right)
\label{eq2},
\end{eqnarray}
the electric field \textbf{E} and the magnetic field \textbf{B} are defined as
\begin{eqnarray}
\begin{array}{l}
{\bf{E}} = \int {\int {m\sqrt {\frac{{\cos \theta '}}{{\cos \theta }}} } } \left[ {{{\bf{e}}_p}^\prime \left( {{\bf{q}} \cdot {{\bf{e}}_p}} \right) + {{\bf{e}}_s}\left( {{\bf{q}} \cdot {{\bf{e}}_s}} \right)} \right]{e^{i{\bf{k'}} \cdot {\bf{r'}}}}d\Omega \\
{\bf{B}} = \int {\int {m\sqrt {\frac{{\cos \theta '}}{{\cos \theta }}} } } \left[ {{{\bf{e}}_s}\left( {{\bf{q}} \cdot {{\bf{e}}_p}} \right) - {{\bf{e}}_p}^\prime \left( {{\bf{q}} \cdot {{\bf{e}}_s}} \right)} \right]{e^{i{\bf{k'}} \cdot {\bf{r'}}}}d\Omega
\end{array}
\label{eq3}
\end{eqnarray}
where $d\Omega = \sin \theta 'd\theta 'd\varphi$, $\bf{k}'$ and $\bf{r}'$ are the wave vector and the position vector of a target point in image space, $\theta '$ is the intersection angle between $\bf{k}'$ and $z$-axis. The relationship between the $\theta$ and $\theta '$ is given by Abbe's sine condition, $\sin \theta = m\sin \theta '$, $m$ is the magnification of the imaging system. The wave vector $\bf{k}'$ and unit vector ${{\bf{e}}_p}^\prime$ can be define as $\bf{k}' = 2\pi /\lambda \left( { - \sin \theta '\cos \varphi , - \sin \theta '\sin \varphi , - \cos \theta '} \right)$, ${{\bf{e}}_p}^\prime  = \left( {\cos \theta '\cos \varphi ,\cos \theta '\sin \varphi , - \sin \theta '} \right)$, respectively. We can theoretically calculate the intensity and polarization distributions of scattered light of a single nanoparticle in the image and Fourier space according to the Eq. (1)- (3).

\begin{figure}
\centering
\includegraphics[width=3.2in]{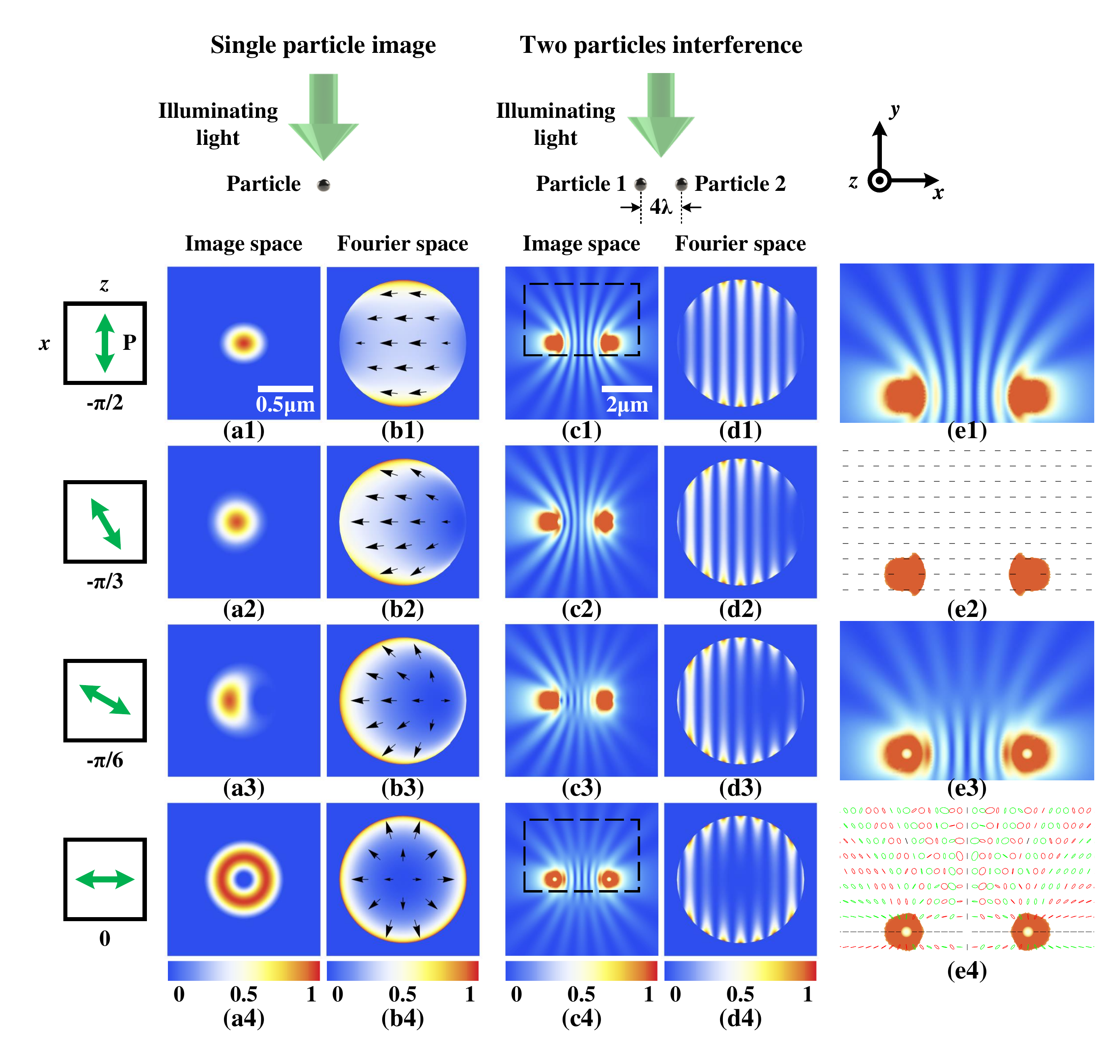}
\caption{Theoretical calculation of the intensity and polarization distributions of the scattered light. The patterns of the scattered light of single nanoparticle in the image space are shown in (a1)-(a4) and the patterns in the Fourier space are shown in (b1)-(b4) with the polarization of the illuminating light in $xz$-plane. The interference pattern of the scattered light of the two nanoparticles in the image space are shown in (c1)-(c4) and the patterns in the Fourier space are shown in (d1)-(d4) with the different polarization of the illuminating light. (e1) and (e2) ((e3) and (e4)) are the intensity and polarization distribution of the partial area (black dashed box) in (c1) ((c4)). The black lines, red and green ellipses represent the linear polarization, right and left-handed elliptic polarization, respectively.}
\label{fig2}
\end{figure}

Here we consider three cases of the polarization of the illuminating light along $x$- and $z$-axis, which correspond to the dipole orientations of \textbf{Case 1}: $(\alpha, \beta)=(- \pi/2,0)$, \textbf{Case 2}: $(\alpha, \beta)=(0,0)$, and \textbf{Case 3} as the intermediate cases between case 1 and case 2: $(\alpha, \beta)=(-\pi/2<\alpha<0$ and $0<\alpha<\pi/2,0)$, respectively. The intensity and polarization distributions of the scattered light are shown in Figs. 2(a1)-(a4) and 2(b1)-(b4). The intensity distribution has a symmetric elliptical profile in the image space (Fig. 2(a1)) and saddle-like shape in the Fourier space (Fig. 2(b1)) for the case 1. Simultaneously, the polarization in the image and Fourier space is linear, aligned along $x$-axis. Case 3 for the arbitrary polarization of the illuminating light is discussed in our previous work \cite{Jin21}. The dark area in the image and Fourier space will appear on the right and gradually move towards the center when the polarization of the illuminating light rotates from $-\pi/2$ to $0$ (Figs. 2(a2)-(b2) for $-\pi/3$ and Figs. 2(a3)-(b3) for $-\pi/6$), and the polarization is also changed. For the symmetry, the dark area in the image and Fourier space will gradually move from the center to the left when the polarization of the illuminating light rotates from $0$ to $\pi/2$. In contrast, for the case 2, the intensity distribution profile of the scattered light changes into a donut hole in the image space (Fig. 2(a4)) and Fourier space (Fig. 2(b4)), while polarization distribution shows a polarization vortex with radial polarization.

Now we consider the interference of the scattered light by two nanoparticles. We assume that the two trapped nanoparticles are aligned along $x$-axis, i.e. $x_1=-2\lambda$, $x_2=2\lambda$, as shown in Fig. 2(b). We calculate the intensity distribution of the scattered light interference of the two dipoles in the Fourier space. The interference intensity distribution of the two linearly polarized light fields can be written as
\begin{eqnarray}
I_{{\mathop{\rm int}} }^F\left( {{x^F},{y^F}} \right) = I_1^F + I_2^F + 2\sqrt {I_1^FI_2^F} \cos \phi \cos \eta
\label{eq4},
\end{eqnarray}
where $I_j^F$ is the intensity of the scattered light of single nanoparticle in the Fourier space, $j = 1, 2$ correspond to the two nanoparticles, $\phi$ is the phase difference and $\eta$ is the polarization intersection angle between the two light fields at same point. The $\phi$ and $\eta$ are spatial dependent in the Fourier space.

In the image space, we can calculate the interference distribution obtained from Eq. (3). However, the calculation is very complex. Here we adopt the paraxial approximation for the dipole scattering (regard as two spot light source) to obtain the simple expressions as below. The interference distribution in image space is determined only by $E_x$ and $E_y$ (paraxial approximation; $E_z$ negligible). The complex amplitude form of the electric field at the image space (x,y) can be written as ${{\bf{E}}_j}(x,y) = \frac{1}{{{r_j}}}{{\bf{p}}_j}{e^{ik{r_j}}}$, where $r_j$ is the distance between the field (x,y) and the position of the nanoparticle in the image space. The polarization of the scattered light fields ${{\bf{p}}_j} = \left( {{p_{xj}},{p_{yj}}} \right)$ can be resolved into x and y components. Thus the superposition of the two light fields in the image space (x,y) can be written as ${E_x} = \frac{1}{{{r_1}}}{p_{x1}}{e^{ik{r_1}}} + \frac{1}{{{r_2}}}{p_{x2}}{e^{ik{r_2}}}$, ${E_y} = \frac{1}{{{r_1}}}{p_{y1}}{e^{ik{r_1}}} + \frac{1}{{{r_2}}}{p_{y2}}{e^{ik{r_2}}}$, where $\left( {{p_{xj}},{p_{yj}}} \right) = \left( {\frac{{x - {x_j}}}{{{r_j}}}\cos \alpha  + \sin \alpha ,\frac{y}{{{r_j}}}\cos \alpha } \right)$. The intensity and polarization distributions of the interference of two scattered light fields need to be considered simultaneously. According to the Stokes parameters, ${S_0} = {\left| {{E_x}} \right|^2} + {\left| {{E_y}} \right|^2}$, ${S_1} = {\left| {{E_x}} \right|^2} - {\left| {{E_y}} \right|^2}$, ${S_2} = {E_x}E_y^* + {E_y}E_x^*$, ${S_3} = i\left( {{E_x}E_y^* - {E_y}E_x^*} \right)$, we can calculate the intensity distribution of interference and give the polarization distribution in the image space.

We plot the intensity distributions with the interference fringes in the image and Fourier space for the three cases as shown in Figs. 2(c1)-(c4) and (d1)-(d4). For case 1, the interference fringes in the image space presents hyperbolic-like lineshape (Figs. 2(c1)) and in the Fourier space present the vertical stripes (Figs. 2(d1)) which are superimposed on the original saddle-like intensity distribution. The polarization distribution in the image space (Fig. 2(e2)) and Fourier space show the homogeneous linear polarization along $x$-axis. In comparison, the intensity distributions in the image and Fourier space for the case 2 are shown in Figs. 2(c4) and (d4), respectively. The interference fringes in the Fourier space also presents the vertical stripes which are superimposed on the original doughnut intensity distribution. However, the interference fringes in the image space are quite different from the case 1, and show the $\pi$ shift of interference fringes between inside and outside the circular demarcation line, where the circular demarcation line is defined as the half of the distance between the two nanoparticles from the center of the two nanoparticles. On the circular demarcation line, the polarization of scattered light of two nanoparticles is orthogonal. The polarization intersection angle is less than $\pi/2$ outside the circular demarcation line (far region). However, the polarization intersection angle is more than $\pi/2$ inside the circular demarcation line (center region). So the polarization of scattered light of two nanoparticles in the center region almost is aligned the opposite direction, which looks like that a $\pi$ global phase is introduced in the center region. This induces relative $\pi$ shift of interference fringes between inside and outside the circular demarcation line. Therefore, the bright and dark interference fringes in the central region between the two nanoparticles in the image space are inverse and almost same in the far region as case 1 compares with the case 2. Moreover, the polarization distribution in the image space presents the periodic variations of the ellipticity of polarization as shown in Fig. 2(e4). The interference fringes for Case 3 are studied as shown in Figs. 2(c2)-(c3) and (d2)-(d3) for $-\pi/3$ and $-\pi/6$ respectively. The interference fringes in the image and Fourier space are also superimposed on the corresponding intensity distribution of the single nanoparticle scattering (Fig. 2(a2)-(b2) and (a3)-(b3). The bright and dark interference fringes in the central region between the two nanoparticles for the image space are gradually shifted to inverse state. These results show the distinct features of the interference pattern for the scattered vector light fields.

\section{Results}

\subsection{Rotation of  the input polarization of illuminating light}

\begin{figure}
\centering
\includegraphics[width=3.2in]{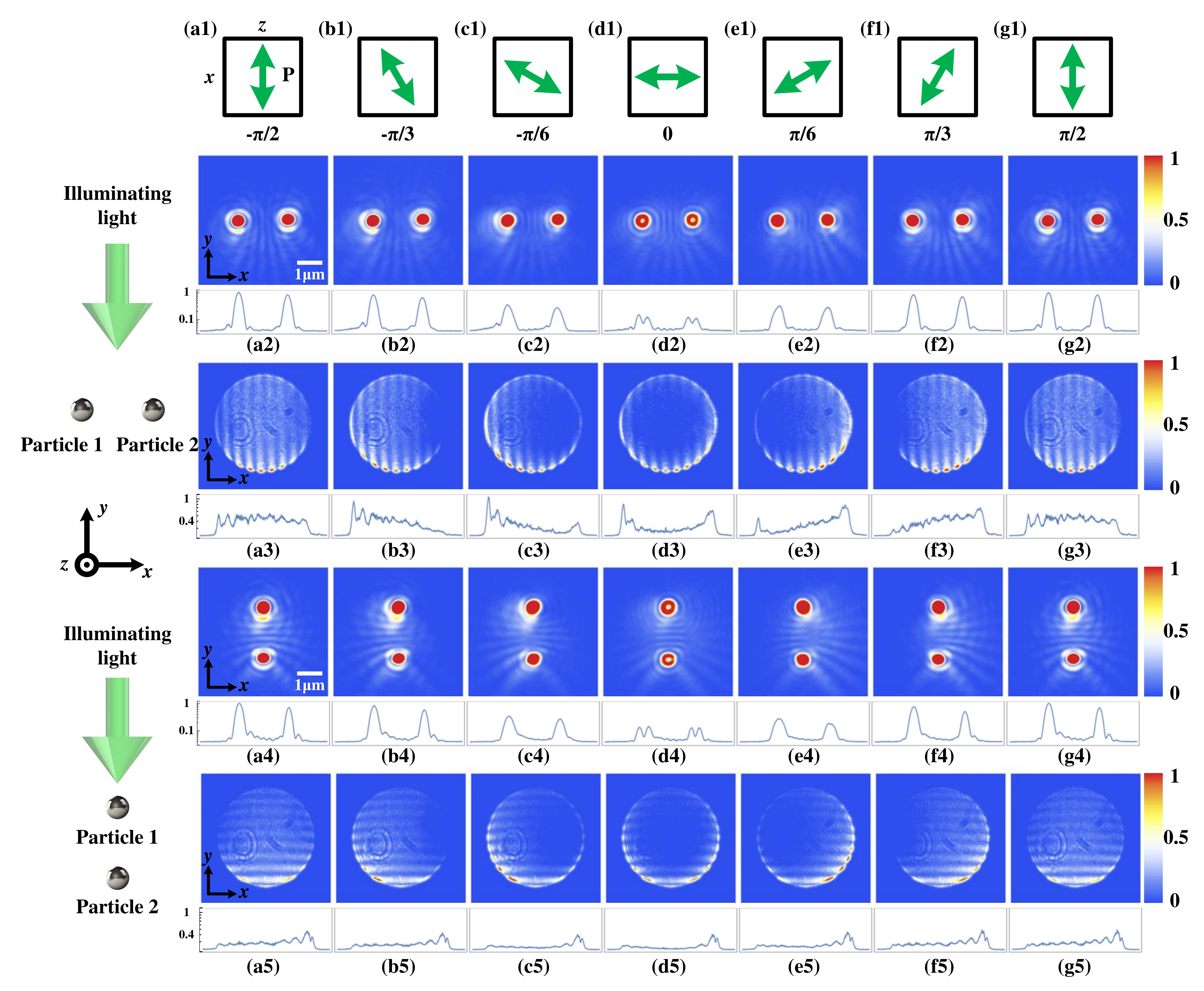}
\caption{The interference image of the scattered light from the two nanoparticles versus the linear polarization of the illuminating light in the image and Fourier space. (a1)-(g1) The polarization of the illuminating light in $xz$-plane. (a2)-(g2) and (a3)-(g3) are the measured results in the image space ($xy$-plane) and Fourier space ($xy$-plane) in the case of the two nanoparticles along $x$-axis. (a4)-(g4) and (a5)-(g5) are the measured results in the image space ($xy$-plane) and Fourier space ($xy$-plane) in the case of the two nanoparticles along $y$-axis.The line plots below each image of (a2)-(g2) and (a4)-(g4) is the intensity distribution from cross-sectional view through the center of the two nanoparticles. The line plot below each image of (a3)-(g3) and (a5)-(g5) is the intensity distribution from the center cross-sectional view of the images alone the direction of the two nanoparticles. A few specks in (a3)-(g3) and (a5)-(g5) is caused by some dusts on the microscope objective. The Rayleigh diffraction rings centering on the position of nanoparticles caused by the diffraction of the nanaparticle are also observed in the images of the image space (a2)-(g2) and (a4)-(g4).}
\label{fig3}
\end{figure}

Now we give the experimental results on the interference of the scattered light of two nanoparticles in the image and Fourier space. Firstly, we discuss the interference pattern of the light scattered by the two nanoparticles along the $x$-axis for the different input linear polarization directions of the illuminating light as shown in Fig. 3. When the linear polarization of the illuminating light is aligned along the $x$-axis (orthogonal to optical axis of the objective lens, as shown in Fig. 3(a1)), i.e., $\alpha=-\pi/2$, the interference pattern in the image space shows several interference fringes, which present hyperbolic-like lineshape and become shallower close to the middle as shown in Fig. 3(a2). In parallel, the interference pattern in the Fourier space is similar with the double-pinhole interference \cite{Norrman20,Casta2021} as shown in Fig. 3(a3), which is consistent with the theoretical simulation of  Fig. 2(d1). When the polarization of the illuminating light is rotated anti-clockwise in the $xz$-plane by turning the HW3, the images of the scattered light for every $\pi /6$ radians of the polarization angle are recorded in the image space (Figs. 3(a2)-(g2)) and Fourier space (Figs. 3(a3)-(g3)) respectively. The total intensity of the scattered light gradually becomes weaker (stronger) as we scan the polarization from $-\pi/2$ to $0$ ($0$ to $\pi/2$), which can be seen from the line plots in Fig. 3(a2)-(g2). The bright and dark interference fringes in the central region in the image space are gradually shifted to inverse state as shown in Figs. 3(b3) and (c3) for the polarization values of $-\pi/3$ and $-\pi/6$ and Fig. 3(e3) and (f3) for $\pi/6$ and $\pi/3$ because of the change of polarization distribution. When the dipole orientation is aligned along $z$-axis, the scattered light represents a polarization vortex field with radial polarization (the intensity distribution profile of scattered light presents a donut hole) and the interference fringes in the image space show the special feature with the $\pi$ shift of the interference fringes (Fig. 3(d2)) in the central region, which is in good agreement with the theoretical prediction.

Then we change the relative positions of two trapped nanoparticles to along $y$-axis by titling the propagation directions of the two trapping beams to shift the focus point in the radial direction. Fig3. (a4)-(g4) and (a5)-(g5) are the measured results in the image and Fourier space in the case of the two nanoparticles along $y$-axis, respectively. The interference patterns are similar with the case of the two nanoparticles along $x$-axis. Note that in our experiment, the intensity of the lower part of the images is stronger than the intensity of upper part, which means that the scattered light in forward direction of the illuminating light ($y$-axis negative direction) is a little stronger than the backward direction ($y$-axis positive direction), which can be seen in Fig. 3.

\subsection{Change distance between two nanoparticles}
\begin{figure}
\centering
\includegraphics[width=3.2in]{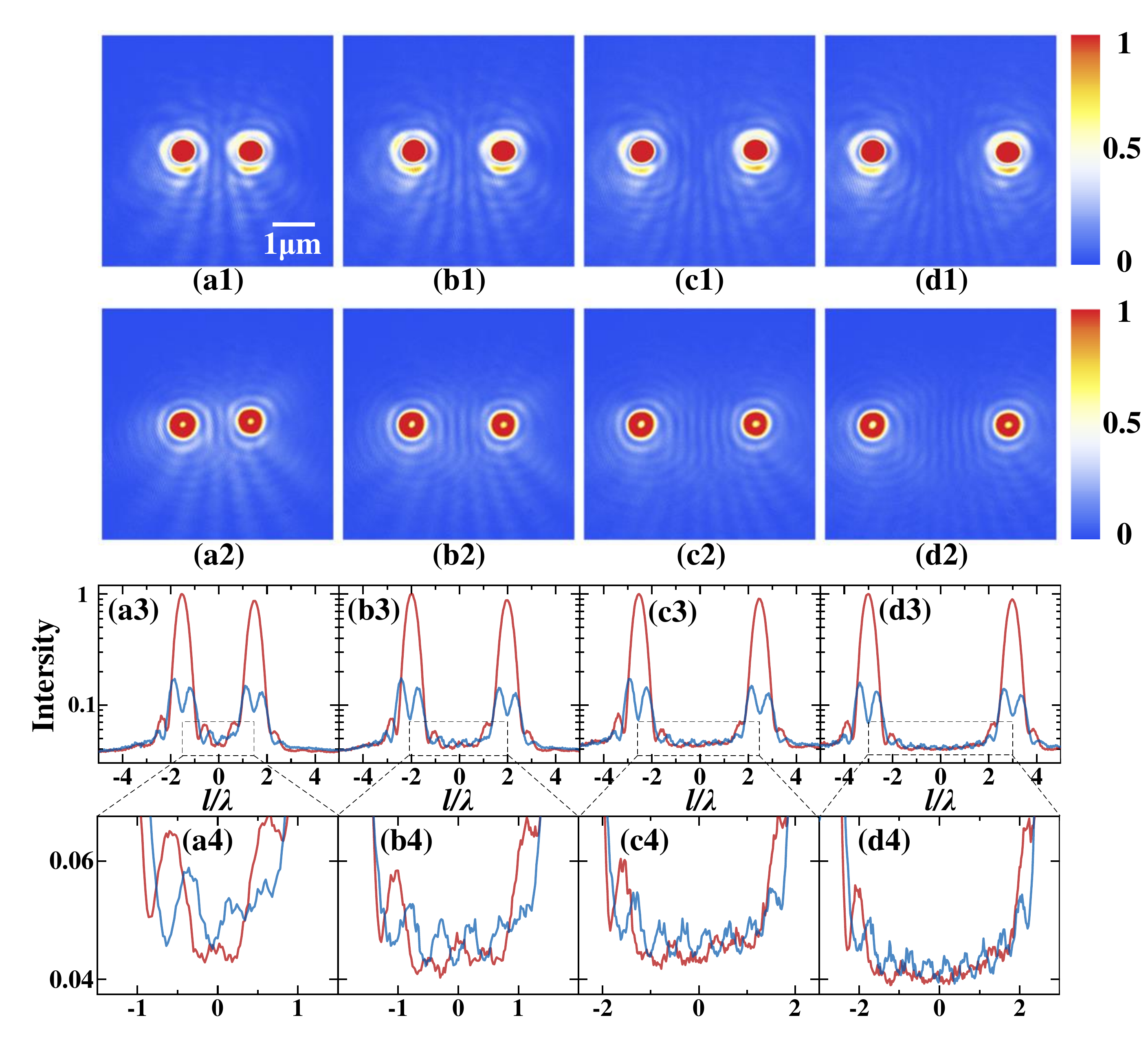}
\caption{The interference images of scattered light versus the distance between the two nanoparticles in the image space. (a1)-(d1) and (a2)-(d2) are the measured results at different distances with the polarization of illuminating light along $x$- and $z$-axis respectively. (a3)-(d3) are the intensity distributions on the connecting line of the two nanoparticles. (a4)-(d4) are the corresponding partial enlargement (black dashed boxes) of (a3)-(d3).}
\label{fig4}
\end{figure}

Measuring the position of a dipolar scatter in a focus light field was studied theoretically \cite{PhysRevA.100.043821}. Here, we give a new method to determine the exact distance between the two nanoparticles by using the interference fringes of the scattered light. In the experiment, we carefully control the distance between the two nanoparticles and record the interference fringes pattern for the case 1 and 2. The experimental results are plotted in Fig. 4 (Figs. 4(a1)-(d1) for the case 1 and Figs. 4(a2)-(d2) for the case 2). We can see the different number of the interference fringes for the different distance between the two nanoparticles. Furthermore, we plot the intensity distributions at the line connecting the positions of the two nanoparticles on the images for the two cases as shown in Figs. 4(a3)-(d3). The red and blue lines correspond to the case 1 and 2, respectively. Figs. 4(a4)-(d4) are the corresponding partial enlargement (black dashed boxes) of Figs. 4(a3)-(d3), in which we can clearly see the bright and dark interference fringes are inverse for the two different case 1 and 2 due to the $\pi$ shift of the interference fringes. Moreover, we measure and control the distance between the two nanoparticles. Every pixel of our camera is 6.5 $\mu$m. The mean distance of the two adjacent bring (or dark) fringes is 23.85 pixel in the image of our experimental results, which equal to half of the wavelength of the illuminating light (532/2 nm) in the actual space. This is a very accurate ruler to determine the spatial distance. Thus, the magnification of our imaging system is about 583 times. Then we can control and calibrate the distance between the two nanoparticles by measuring the distance of the images of two nanoparticles. It can be obtained that the actual distance of the two nanoparticles are $d = 3.02 \lambda, 3.97 \lambda, 5.02 \lambda, 5.99 \lambda$ from Fig. 4(a1)-(d1). The precision is limited by the size of the camera pixel and the magnification of the imaging system. It can be reached to 10 nm for our system.

\section{Conclusion}

We have experimentally observed the interference of two scattered vector fields from two optically levitated nanoparticles in vacuum. We employ the distinguished features of this system including an environment free of the particle-substrate interactions, providing a dark background and high signal-noise ratio for detecting the dipole scattering, producing the scattered vector light fields. The distinct interference patterns in intensity and polarization distributions are observed for the different dipole orientations. Especially, the interference fringes in the image space for two scattered vector light fields show the $\pi$ shift of the interference fringes in the central region between the two nanoparticles, in which the bright and dark interference fringes are inverse and almost same in the far region for the case 1 and case 2. Furthermore, this work provides a precise tool to measure the spatial scale of the optically levitated system.

\appendix

\section{Appendixes}
\textbf{Intensity distribution of the scattered light of nanoparticle in the image space.}
The intensity distribution of the scattered light in the image space is given by the Eq. (3). The position vector of a target point in focus plane is ${\bf{r'}} = \left( {\rho \cos \varphi ',\rho \sin \varphi ',z'} \right)$. Thus, we can write the x and y components of electric field \textbf{E} as
\begin{widetext}
\begin{eqnarray}
\begin{array}{l}
\left( {\begin{array}{*{20}{c}}
{{E_x}}\\
{{E_y}}
\end{array}} \right) = \int {\int {m\sqrt {\frac{{\cos \theta '}}{{\cos \theta }}} } } {e^{ - 2i\pi /\lambda \left( {\rho \sin \theta '\cos \left( {\varphi  - \varphi '} \right) + z'\cos \theta '} \right)}}\sin \theta 'd\theta 'd\varphi  \times \\
\left\{ {\left( {\begin{array}{*{20}{c}}
{\cos \theta '\cos \varphi }\\
{\cos \theta '\sin \varphi }
\end{array}} \right)\left[ {\sin \alpha \cos \theta \cos \left( {\varphi  - \beta } \right) + \cos \alpha \sin \theta } \right] + \left( {\begin{array}{*{20}{c}}
{\sin \varphi }\\
{ - \cos \varphi }
\end{array}} \right)\sin \alpha \sin \left( {\varphi  - \beta } \right)} \right\}
\end{array},
\label{eq4}
\end{eqnarray}
\end{widetext}
and the x and y components of magnetic field \textbf{B} as
\begin{widetext}
\begin{eqnarray}
\begin{array}{l}
\left( {\begin{array}{*{20}{c}}
{{B_x}}\\
{{B_y}}
\end{array}} \right) = \int {\int {m\sqrt {\frac{{\cos \theta '}}{{\cos \theta }}} } } {e^{ - 2i\pi /\lambda \left( {\rho \sin \theta '\cos \left( {\varphi  - \varphi '} \right) + z'\cos \theta '} \right)}}\sin \theta 'd\theta 'd\varphi  \times \\
\left\{ {\left( {\begin{array}{*{20}{c}}
{\sin \varphi }\\
{ - \cos \varphi }
\end{array}} \right)\left[ {\sin \alpha \cos \theta \cos \left( {\varphi  - \beta } \right) + \cos \alpha \sin \theta } \right] - \left( {\begin{array}{*{20}{c}}
{\cos \theta '\cos \varphi }\\
{\cos \theta '\sin \varphi }
\end{array}} \right)\sin \alpha \sin \left( {\varphi  - \beta } \right)} \right\}
\end{array}.
\label{eq5}
\end{eqnarray}
\end{widetext}

\begin{acknowledgments}
This research was supported by the National Natural Science Foundation of China (Grant Nos. 12034011 and 61975101), the Shanxi “1331 Project” Key Subjects Construction, and the Xplorer Prize.
\end{acknowledgments}

\nocite{*}

\bibliography{interference}
\bibliographystyle{apsrev4-1}

\end{document}